\documentclass[preprint,11pt]{elsarticle}

\usepackage{fancyhdr}
\usepackage{fullpage}
\usepackage{amsmath, mathtools}
\usepackage{amsbsy}
\usepackage{amssymb}
\usepackage{amscd}
\usepackage{amsfonts}
\usepackage{supertabular}
\usepackage{tablefootnote}
\usepackage{graphics}
\usepackage{verbatim}
\usepackage{epsfig}
\usepackage{xspace}
\usepackage{euscript}
\usepackage{alltt}
\usepackage{boxedminipage}
\usepackage{float}
\usepackage[colorlinks]{hyperref}
\usepackage{cleveref}
\usepackage{color}
\usepackage[all]{xy}
\usepackage{t1enc}
\usepackage{times,exscale}
\usepackage{graphicx,calc}
\usepackage{subfig}
\usepackage[ruled,vlined]{algorithm2e}
\biboptions{sort&compress}
\usepackage{epstopdf}
\usepackage{array,multirow}

\usepackage{graphicx}		        
\usepackage{amsmath,amssymb}		
\usepackage{subfig}					
\usepackage{multirow}
\usepackage{booktabs}

\usepackage{comment}

\biboptions{sort&compress}
\journal{Computer Physics Communications}

\newcommand{\iu}{i}

\DeclareMathOperator{\Hm}{\mathbf{H}}
\DeclareMathOperator{\M}{\mathbf{M}}
\DeclareMathOperator{\LL}{\mathbf{L}}
\DeclareMathOperator{\br}{\mathbf{\widetilde{R}}}

\DeclareMathOperator{\bvh}{\mathbf{\widetilde{V}_{H}}}
\DeclareMathOperator{\bvxc}{\mathbf{V_{xc}}}

\DeclareMathOperator{\bvxcm}{\mathbf{V_{xc}^{mGGA}}}
\DeclareMathOperator{\bvxcoep}{\mathbf{V_{xc}^{OEP}}}
\DeclareMathOperator{\bvxhf}{\mathbf{V_{x}^{HF}}}
\DeclareMathOperator{\bY}{\mathbf{Y}}
\DeclareMathOperator{\bD}{\mathbf{D}}
\DeclareMathOperator{\bP}{\mathbf{P}}
\DeclareMathOperator{\bvnl}{\mathbf{{V}_{NL}}}

\DeclareMathOperator{\F}{\mathbf{F}}

\DeclareMathOperator{\chizero}{\mathbf{X_{00}}}
\DeclareMathOperator{\blambdaxhf}{\mathbf{\Lambda_{x}^{HF}}}
\DeclareMathOperator{\blambdacrpa}{\mathbf{\Lambda_{c}^{RPA}}}
\DeclareMathOperator{\bvxclgm}{\mathbf{V_{xc}^{LDA/GGA/mGGA}}}
\DeclareMathOperator{\bvxclgmm}{\mathbf{V_{xc}^{LDA/GGA}}}
\journal{arXiv}

\begin{document}

\begin{frontmatter}

    \title{SPARC-atomSFE: Spectral finite-element package for atomic structure calculations in density functional theory}

    \author[gatech_coc]{Qihao Cheng}
    \author[gatech_coe]{Shubhang Krishnakant Trivedi}
    \author[gatech_coc,gatech_coe]{Phanish Suryanarayana\corref{cor}}

    \address[gatech_coc]{College of Computing, Georgia Institute of Technology, Atlanta, GA 30332, USA}
    \address[gatech_coe]{College of Engineering, Georgia Institute of Technology, Atlanta, GA 30332, USA}

    \cortext[cor]{Corresponding Author (phanish.suryanarayana@ce.gatech.edu)}

    \begin{abstract}
We present SPARC-atomSFE, a spectral finite-element package for  accurate and efficient atomic structure calculations within the framework of Kohn-Sham density functional theory. The package supports both all-electron and norm-conserving pseudopotential calculations across a comprehensive hierarchy of exchange-correlation approximations, spanning local, semilocal, and nonlocal functionals.  The latter includes hybrid functionals and the many-body random phase approximation, for which we implement both the generalized Kohn-Sham approach and the optimized effective potential (OEP) method, with OEP necessary for eigenvalue-dependent functionals. Spatial discretization is based on an adaptive grid with element nodes distributed according to the Legendre--Gauss--Lobatto scheme, high-order $C^{0}$-continuous Lagrange polynomial basis functions, and Gauss--Legendre quadrature for numerical integration. We present 
systematic convergence studies and identify the computational parameters required to 
achieve target accuracies. We validate the accuracy of SPARC-atomSFE through representative calculations 
spanning the various exchange-correlations approximations, obtaining results that 
generally agree with values in the literature to within $1~\mu\text{Ha}$ or better.
    \end{abstract}

    \begin{keyword}
    Atomic structure \sep
    Density functional theory \sep
    Exchange-correlation functionals \sep
        Finite element method \sep
    All-electron calculations \sep
    Norm-conserving pseudopotential calculations \sep    \\
\end{keyword}

\end{frontmatter}

\noindent \textbf{PROGRAM SUMMARY}

\begin{small}
\noindent
{\em Program Title:} SPARC-atomSFE \\
{\em Developer's repository link:} \url{https://github.com/SPARC-X/SPARC-atomSFE} \\
{\em Licensing provisions:} GNU GPLv3 \\
{\em Programming language:} Python \\
{\em Nature of problem:} Solution of the radial Kohn--Sham equations and their generalized counterpart for isolated atoms, across local, semilocal, and nonlocal exchange-correlation functionals, for both all-electron and pseudopotential calculations. \\
{\em Solution method:} Spatial discretization is based on an adaptive grid with element nodes distributed according to the Legendre--Gauss--Lobatto scheme, high-order $C^{0}$-continuous Lagrange polynomial basis functions, and Gauss--Legendre quadrature for numerical integration. The Kohn--Sham equations are solved via fixed-point iteration, i.e., the self-consistent field method, with an additional outer loop for nonlocal exchange-correlation functionals. \\
{\em Additional comments including restrictions and unusual features:} Currently restricted to spin-unpolarized, non-relativistic calculations, with support for a selected set of exchange-correlation functionals. \\
\end{small}

\section{Introduction}
\label{sec:intro}

Over the past several decades, electronic structure calculations based on Kohn--Sham density functional theory (DFT)~\cite{kohn1965self, hohenberg1964inhomogeneous} have become a cornerstone of research in the materials and chemical sciences, owing to the fundamental physical insights they provide and their strong predictive capability. Having its origins in the first principles of quantum mechanics, Kohn--Sham DFT offers an effective balance of conceptual simplicity, broad applicability, and favorable accuracy-to-computational-cost ratio relative to other \textit{ab initio} methods. Its widespread use spans isolated systems, such as molecules and clusters; semi-infinite systems, such as nanotubes and surfaces; and bulk three-dimensional systems.

The accuracy and computational cost of DFT calculations are primarily determined by the choice of exchange--correlation functional, which models the many-body electron interactions and constitutes the main approximation in the Kohn--Sham formalism. Since the universal exchange--correlation functional is unknown, a hierarchy of approximations has been developed and organized within Jacob's ladder~\cite{jacobladderperdew}, with successive rungs generally providing improved accuracy at increased computational cost. The lowest four rungs, which include local, semilocal, and hybrid functionals, are the most widely used in practice. The fifth and highest rung consists of nonlocal many-body correlation methods, with the random phase approximation (RPA) serving as a representative example~\cite{ren2012random, eshuis2012electron}. Fifth-rung functionals such as RPA can provide benchmark-level accuracy~\cite{RenRPA3, KresseRPAlattice, Hutterliquidwater, JiangRPAstability, Tkatchenkovanderwaals, ThygesenAdsorptionenergies, DierRPA, RPA-SELF-CONSISTENT-PBE0-PITTS-PSEUDOPOTENTIAL-SOLIDS}, but at computational costs that are orders of magnitude higher than those of lower-rung functionals~\cite{KresseRPAForces, ShikharRPA, boqinRPA}. This high cost has hindered their systematic testing and development, limited their practical use, particularly in self-consistent calculations requiring the optimized effective potential (OEP) method~\cite{oepjdtalman, Sham-Schluter-equation-OEP}, and complicated the generation of high-quality training data for machine-learned models.

Atomic structure calculations~\cite{lehtola2019review} --- electronic 
structure problem is solved in radial coordinates by exploiting the spherical symmetry 
of isolated atoms --- form an integral component of the DFT infrastructure and provide 
an attractive setting for the assessment and development of exchange--correlation 
functionals. In particular, they are used to generate pseudopotentials~\cite{hamann2013optimized, 
troullier1991efficient, shojaei2023soft}, atom-centered bases for the projector 
augmented wave (PAW) method~\cite{blochl1994projector, holzwarth2001projector}, and 
inputs for Hubbard-corrected DFT (DFT+U)~\cite{anisimov1991density, anisimov1991band, bhowmik2025real}. The orbitals, densities, and potentials so obtained are further used to construct 
initial guesses that accelerate SCF convergence~\cite{lehtola2019assessment, 
lehtola2020efficient, xu2021sparc}; to form the numerical atomic orbitals (NAOs) 
employed in linear-combination-of-atomic-orbitals (LCAO) 
methods~\cite{CONQUEST, garcia2020siesta, GPAW, blum2009ab}; and to facilitate the 
analysis of DFT results such as the projected density of states 
(PDOS)~\cite{herath2020pyprocar, wang2021vaspkit}. For exchange--correlation 
development, high-quality reference data from coupled-cluster 
[CCSD(T)]~\cite{CCSD(T)inversion, CCSD(T)openshellatoms}, configuration-interaction 
(CI)~\cite{chakravarthyhelium, chakravarthyberylliumandothers}, and quantum Monte 
Carlo (QMC)~\cite{umrigargonze} calculations are readily available for atoms, 
enabling rigorous and systematic benchmarking. Furthermore, because atomic structure 
calculations are orders of magnitude more efficient than their 3D 
counterparts, they enable all-electron treatments across the entire periodic table, 
providing a valuable platform for generating high-fidelity training data for 
machine-learned exchange--correlation models~\cite{realspacesecondorderML, MLCI, trivedi2026spectral} and 
for constructing descriptors for machine-learned force fields 
(MLFFs)~\cite{lei2022universal, timmerman2024overcoming, qiao2022informing}.

A number of solution strategies have been developed for atomic structure calculations within the framework of Kohn-Sham DFT~\cite{holzwarth2001projector, vcertik2013dftatom, holzwarth2022cubic, AndraeExamination2001, lehtola2023meta, vcertik2024high, lehtola2020fully, romanowski2007b, romanowski2009adaptive, lehtola2019fully, cinal2020highly, fischerHFatoms, OZAKI20111245, secondorderKSMP2, gwasphericalatomshellgren, vacondiopaper, scRPAgorlingpaper, trivedi2026spectral, hamann2013optimized, bhowmik2025spectral, fuchs1999ab, OLIVEIRA2008524, mGGA_USPP, PBE0_RRKJ_psp, RSH_NC_psp}, including shooting-type methods~\cite{hamann2013optimized, fuchs1999ab, holzwarth2001projector, vcertik2013dftatom, OLIVEIRA2008524}, spline approaches~\cite{holzwarth2022cubic, gwasphericalatomshellgren, vacondiopaper}, finite-difference methods~\cite{AndraeExamination2001}, finite-element methods~\cite{lehtola2019fully, lehtola2020fully, lehtola2023meta, OZAKI20111245, romanowski2007b, romanowski2009adaptive, vcertik2024high, trivedi2026spectral}, spectral schemes~\cite{bhowmik2025spectral}, pseudospectral schemes~\cite{cinal2020highly}, and Gaussian basis set approaches~\cite{scRPAgorlingpaper}. Most of these solvers generally focus on the local density approximation (LDA) \cite{kohn1965self} and Generalized Gradient Approximation (GGA) \cite{perdew1996generalized} to the exchange-correlation \cite{bhowmik2025spectral, OZAKI20111245, lehtola2020fully, AndraeExamination2001, vcertik2013dftatom,vcertik2024high, romanowski2007b, romanowski2009adaptive, holzwarth2001projector, fuchs1999ab, OLIVEIRA2008524, hamann2013optimized}, 
with some extensions to meta-GGA~\cite{bhowmik2025spectral, lehtola2023meta, mGGA_USPP, holzwarth2022cubic}, Hartree-Fock (HF) \cite{OZAKI20111245, lehtola2020fully, AndraeExamination2001, lehtola2019fully, cinal2020highly, fischerHFatoms}, and hybrid~\cite{bhowmik2025spectral, PBE0_RRKJ_psp, RSH_NC_psp} functionals, while support for fifth-rung functionals such as RPA and MP2--- which are eigenvalue-dependent and therefore require the OEP method --- remains an active area of development \cite{secondorderKSMP2, gwasphericalatomshellgren, vacondiopaper, 
scRPAgorlingpaper, trivedi2026spectral}.   In particular, solvers are typically tailored toward either all-electron or pseudopotential calculations, with support for fractional occupations, charged atoms, various exchange-correlation functionals, and the OEP method varying considerably across codes. Moreover, many existing implementations are in compiled languages such as Fortran, which can make interfacing with modern machine learning frameworks less straightforward, and not all codes are openly available. An atomic solver spanning the full exchange--correlation hierarchy --- from local and semilocal to nonlocal and many-body --- within a single, open-source, Python-based framework, for both all-electron and pseudopotential calculations, is the goal of the present effort.

In this work, we present SPARC-atomSFE, an open-source, Python-based spectral finite-element atomic structure code that supports both all-electron and norm-conserving pseudopotential calculations across the full hierarchy of exchange--correlation approximations, from local and semilocal functionals to hybrid and many-body fifth-rung functionals. We implement both the generalized Kohn--Sham and 
OEP approaches for nonlocal potentials, with the OEP method required for 
eigenvalue-dependent functionals in the fifth rung. The package also supports 
fractional orbital occupations and charged atoms. We demonstrate the accuracy of the package through systematic convergence studies and validation against literature results for a range of exchange-correlation functionals.

The remainder of this manuscript is organized as follows. In 
Section~\ref{Sec:Equations}, we present the mathematical formulation of atomic 
structure calculations within Kohn--Sham DFT. In Section~\ref{Sec:FEM}, we describe 
the spectral finite-element framework for such calculations. In Section~\ref{Sec:Implementation}, we discuss the 
implementation of the framework in SPARC-atomSFE. In Section~\ref{Sec:Results}, we assess 
the accuracy and performance of the code. Finally, we offer concluding remarks in 
Section~\ref{Sec:conclusions}.

\section{Atomic structure problem}
\label{Sec:Equations}

Consider an isolated atom with atomic number $Z$ and $N_e$ electrons. Exploiting spherical symmetry and neglecting spin polarization, the generalized Kohn--Sham DFT energy functional in radial coordinates takes the form \cite{vcertik2024high,bhowmik2025spectral, trivedi2026spectral}:
\begin{align}
    E[\widetilde{R}_{nl},\lambda_{nl}] = T_s[\widetilde{R}_{nl}] + E_{xc}[\rho, \nabla\rho, \tau, \widetilde{R}_{nl},\lambda_{nl}] + E_{H}[\rho] + E_{nuc}[\rho] \,, \label{Eq:energy}
\end{align}
where $T_s$ is the electronic kinetic energy; $E_{xc}$ is the exchange-correlation energy, which can be split into exchange ($E_x$) and correlation ($E_c$) components; $E_{H}$ is the Hartree energy corresponding to Coulomb interaction between the electrons; and $E_{nuc}$ is the  energy corresponding to the Coulomb interaction between the nucleus and electrons; $n$, $l$, and $m$ are the principal, azimuthal, and magnetic quantum numbers, respectively; $\widetilde{R}_{nl}/r$ are the radial components of the Kohn--Sham orbitals, with corresponding eigenvalues $\lambda_{nl}$ and occupations $g_{nl} = 2(2l+1)f_{nl}$, where $f_{nl} \in [0,1]$ and $\sum_{nl} g_{nl} = N_e$; $\rho$ is the electron density:
\begin{align} \label{density}
    \rho = \frac{1}{4\pi r^2}\sum_{nl} g_{nl} (\widetilde{R}_{nl})^2 \,; 
\end{align}
$\nabla \rho$ is the density gradient, which in radial coordinates reduces to $\nabla\rho = d\rho/dr$; and $\tau$ is the positive-definite kinetic energy density: $   \tau = \frac{1}{2}\sum_{nl}\frac{g_{nl}}{4\pi}\left[ \left( \frac{1}{r}\frac{d \widetilde{R}_{nl}}{d r} - \frac{\widetilde{R}_{nl}}{r^2}\right)^2 + \frac{l(l+1)}{r^4}(\widetilde{R}_{nl})^2\right]$. The electronic kinetic energy and electrostatic energy components take the form \cite{bhowmik2025spectral, trivedi2026spectral}:
\begin{align}
    T_{s}[\widetilde{R}_{nl}] &= \frac{1}{2} \sum_{nl} g_{nl} \int \left[  \left( \frac{d \widetilde{R}_{nl}(r)}{d r}\right)^2 + \widetilde{R}_{nl}(r) \frac{l(l+1)}{r^2} \widetilde{R}_{nl}(r) \right] d r \,, \label{KE_eq} \\
    E_{H}[\rho] &= \max_{\widetilde{V}_{H}} \left[ -\frac{1}{2} \int \left( \frac{d \widetilde{V}_{H}(r)}{d r} \right)^2 d r + 4\pi \int \rho(r) \,\widetilde{V}_{H}(r) \, r \, d r \right] \,, \\
    E_{nuc}[\rho] &= 4\pi \int \rho(r) \, V_{nuc}(r) \, r^2 \, d r \,, \label{HartreeEnergy}
\end{align}
where $\widetilde{V}_{H}(r)/r$ is the Hartree potential and $V_{nuc}(r) = -Z/r$ is the  Coulomb potential of the nucleus. For local and semilocal exchange-correlation functionals, namely LDA, GGA, and mGGA, the exchange-correlation energy takes the form:
\begin{subequations}
\begin{align}
    E_{xc}^{\rm LDA}[\rho] &= 4\pi \int \rho(r)\,\varepsilon_{xc}^{\rm LDA}(\rho)\, r^2 \, d r \,, \label{Eq:Exc:LDA} \\
    E_{xc}^{\rm GGA}[\rho, \nabla\rho] &= 4\pi \int \rho(r)\,\varepsilon_{xc}^{\rm GGA}(\rho, \nabla\rho)\, r^2 \, d r \,, \label{Eq:Exc:GGA} \\
    E_{xc}^{\rm mGGA}[\rho, \nabla\rho, \tau] &= 4\pi \int \rho(r)\,\varepsilon_{xc}^{\rm mGGA}(\rho, \nabla\rho, \tau)\, r^2 \, d r \,, \label{Eq:Exc:mGGA}
\end{align}
\end{subequations}
where $\varepsilon_{xc}$ is the exchange-correlation energy per electron. For hybrid functionals, the correlation energy takes a local or semilocal form as above, while the exchange energy includes a fraction of the exact (Fock) exchange energy \cite{cinal2020highly, bhowmik2025spectral, trivedi2026spectral}:
\begin{align}\label{eq:exact-exchange-energy}
    E_x^{\rm HF}[\widetilde{R}_{nl}] = -\frac{1}{4} \sum_{\substack{nl \\ n'l'}} g_{nl} g_{n'l'} \sum_{l'\!'=|l-l'|}^{l+l'} \left[
    \begin{pmatrix}
        l & l' & l'\!' \\
        0 & 0 & 0
    \end{pmatrix}^2
    \iint \widetilde{R}_{nl}(r) \widetilde{R}_{n'l'}(r) \nu_{l'\!'}(r,r') \widetilde{R}_{nl}(r') \widetilde{R}_{n'l'}(r') \, drd r' \right] \,,
\end{align}
where $\begin{pmatrix} l & l' & l'\!' \\ 0 & 0 & 0 \end{pmatrix}$ is the Wigner--3j symbol and $\nu_{l'\!'}$ is the radial Coulomb operator:
\begin{equation}\label{eq:radial-coulomb-operator}
    \nu_{l'\!'}(r,r') = \frac{r^{l'\!'}_{<}}{r^{l'\!'+1}_{>}} \,, \qquad r_{<} = \min(r,r') \,, \quad r_{>} = \max(r,r') \,.
\end{equation}
The exact exchange energy can be equivalently written as:
\begin{align}\label{eq:exact-exchange-energy2}
    E_x^{\rm HF}[\widetilde{R}_{nl}] = -\frac{1}{4} \sum_{\substack{nl \\ n'l'}} g_{nl} g_{n'l'} \sum_{l'\!'=|l-l'|}^{l+l'} \left[
    \begin{pmatrix}
        l & l' & l'\!' \\
        0 & 0 & 0
    \end{pmatrix}^2
    \int \widetilde{R}_{nl}(r) \frac{Y_{nln'l'}^{l'\!'}(r)}{r}\widetilde{R}_{n'l'}(r) \, d r\right] \,,
\end{align}
where $Y_{nln'l'}^{l'\!'}(r)$ satisfies the following differential equation \cite{fischerHFatoms, cinal2020highly}:
\begin{subequations}
\begin{align}
    &\Bigg[\frac{d^2}{dr^2} - \frac{l'\!'(l'\!'+1)}{r^2}\Bigg]Y_{nln'l'}^{l'\!'}(r) = - \frac{(2l'\!'+1)}{r}\widetilde{R}_{nl}(r)\widetilde{R}_{n'l'}(r)\, ,
    \label{eq:Ydiff}\\
    &Y_{nln'l'}^{l'\!'}(r=0)  = 0,\quad \Bigg(\frac{dY_{nln'l'}^{l'\!'}(r)}{dr} + l'\!'\frac{Y_{nln'l'}^{l'\!'}(r)}{r}\Bigg)\Bigg\vert_{r\to \infty}  = 0\, ,
    \label{eq:Yboundary}
\end{align}
\end{subequations}
In so doing, the application of the radial Coulomb operator is formulated as the solution of a differential equation.  

The electronic ground state is determined by the variational problem \cite{bhowmik2025spectral, trivedi2026spectral}:
\begin{align} \label{eq:radial-ortho}
    \min_{\widetilde{R}_{nl}} E[\widetilde{R}_{nl}, \lambda_{nl}] \quad \text{s.t.} \quad \int \widetilde{R}_{nl}(r) \widetilde{R}_{n'l'}(r) \, \mathrm{d} r = \delta_{nn'} \delta_{ll'} \,,
\end{align}
with the corresponding Euler--Lagrange equations and boundary conditions:
\begin{subequations}\label{ground_state}
    \begin{align}
        \left[ \mathcal{H}_{l} \equiv -\frac{1}{2}\frac{d^2}{d r^2} + \frac{l(l+1)}{2 r^2} + V_{nuc} + \frac{\widetilde{V}_{\mathrm{H}}}{r} + \hat{V}_{xc}\right] \widetilde{R}_{nl} = \lambda_{nl}\widetilde{R}_{nl} \,, \label{EL:Eigenproblem} \\
        \widetilde{R}_{nl}(r=0) = 0 \,, \quad \widetilde{R}_{nl}(r \rightarrow \infty) = 0 \,, \label{EL:R-boundary} \\
        \frac{d^2 \widetilde{V}_{\mathrm{H}}(r)}{d r^2} = -4\pi r \rho(r) \,, \quad \widetilde{V}_{\mathrm{H}}(r=0) = 0 \,, \quad \widetilde{V}_{\mathrm{H}}(r \rightarrow \infty) = N_e \,. \label{EL:Poisson}
    \end{align}
\end{subequations}
where $\hat{V}_{xc}$ is the exchange-correlation potential operator and $\mathcal{H}_{l}$ is the radial, angular-momentum-dependent Hamiltonian. For LDA, GGA, and mGGA, the exchange-correlation potential operator takes the form \cite{bhowmik2025spectral}:
\begin{subequations}
\begin{align}
    \hat{V}_{xc}^{\mathrm{LDA}} \widetilde{R}_{nl} & =  \left[  \varepsilon_{xc} + \rho\frac{\partial \varepsilon_{xc}}{\partial \rho}  \right]   \widetilde{R}_{nl} \,, \\
    \hat{V}_{xc}^{\mathrm{GGA}} \widetilde{R}_{nl} &= \hat{V}_{xc}^{\mathrm{LDA}} \widetilde{R}_{nl} + \left[ - \frac{d}{d r}\left( \rho\,\frac{\partial \varepsilon_{xc}}{\partial (\nabla\rho)}\right) - \frac{2\rho}{r}\frac{\partial \varepsilon_{xc}}{\partial (\nabla\rho)} \right] \widetilde{R}_{nl} \,, \\
    \hat{V}_{xc}^{\mathrm{mGGA}} \widetilde{R}_{nl} &= \hat{V}_{xc}^{\mathrm{GGA}} \widetilde{R}_{nl} + \frac{1}{2}\left\{-\frac{d}{d r}\left[\rho \frac{\partial \varepsilon_{xc}}{\partial \tau} \frac{d\widetilde{R}_{nl}}{d r}\right]+\frac{1}{r} \frac{d}{d r}\left[\rho \frac{\partial \varepsilon_{xc}}{\partial \tau}\right] \widetilde{R}_{nl}+\rho \frac{\partial \varepsilon_{xc}}{\partial \tau} \frac{l(l+1)}{r^2} \widetilde{R}_{nl}\right\} \,. \label{eq:vkxc-mgga-continuous}
\end{align}
\end{subequations}
For exact exchange, the potential operator takes the form:
\begin{equation}\label{Vxx}
    \hat{V}_x^{\rm HF} \widetilde{R}_{nl} (r) = -\frac{1}{2} \sum_{n'l'} g_{n'l'} \sum_{l'\!'=|l-l'|}^{l+l'} \left[
    \begin{pmatrix}
            l & l' & l'\!' \\
            0 & 0 & 0
    \end{pmatrix}^2
    \widetilde{R}_{n'l'}(r) \int  \nu_{l'\!'}(r,r') \widetilde{R}_{n'l'}(r') \widetilde{R}_{nl}(r') \, d r' \right] \,,
\end{equation}
which can be rewritten, analogously to the exact exchange energy, as:
\begin{align}\label{Vxx2}
    \hat{V}_x^{\rm HF} \widetilde{R}_{nl} (r) = -\frac{1}{2} \sum_{n'l'} g_{n'l'} \sum_{l'\!'=|l-l'|}^{l+l'} \left[
    \begin{pmatrix}
            l & l' & l'\!' \\
            0 & 0 & 0
    \end{pmatrix}^2
    \widetilde{R}_{n'l'}(r) \frac{Y_{nln'l'}^{l'\!'}(r)}{r} \right] \,.
\end{align}
Overall, the electronic ground state is determined by solving the Kohn--Sham eigenproblems in Eqn.~\eqref{ground_state} self-consistently.


\paragraph{RPA-OEP formalism}

For RPA, the exchange energy is taken as the exact exchange, while the correlation energy takes the form \cite{jiang2007random, trivedi2026spectral}:
\begin{equation}\label{eq:Erpa-trace}
    E_c^{\rm RPA} = \frac{1}{2\pi} \sum_{l'\!'} (2l'\!'+1)
    \int \operatorname{Tr}\!\left[
        \widetilde{\chi}_{0,l'\!'}(\iu\omega)\,\nu_{l'\!'}
        + \log\!\left(I - \widetilde{\chi}_{0,l'\!'}(\iu\omega)\,\nu_{l'\!'}\right)
    \right] d\omega \,,
\end{equation}
where $I$ is the identity operator and $\widetilde{\chi}_{0,l'\!'}(r,r';\iu\omega)$ is the radial density response function at imaginary frequency $\iu\omega$:
\begin{equation}\label{eq:chi0-imag}
    \begin{gathered}
        \widetilde{\chi}_{0,l'\!'}(r,r';\iu\omega)
        = 2 \sum_{nl,\, n'l'}
        \bigl(f_{nl}-f_{n'l'}\bigr)\,
        C_{l'\!';ll'}\,
        D_{nl,\, n'l'}(i\omega)\,
        \widetilde{R}_{nl}(r)\,\widetilde{R}_{n'l'}(r)\,
        \widetilde{R}_{nl}(r')\,\widetilde{R}_{n'l'}(r') \,, \\
        C_{l'\!';ll'} = \frac{(2l+1)(2l'+1)}{2l'\!'+1}
        \begin{pmatrix} l & l' & l'\!' \\ 0 & 0 & 0 \end{pmatrix}^{2} \,, \quad
        D_{nl,\, n'l'}(i\omega) = \frac{\lambda_{nl}-\lambda_{n'l'}}{\omega^{2}+(\lambda_{nl}-\lambda_{n'l'})^{2}} \,.
    \end{gathered}
\end{equation}
Here and henceforth, we will use the notation:
\begin{equation}\label{eq:rpa-operator-convention}
    (AB)(r,r'\!') = \int A(r,r')\,B(r',r'\!')\,d r' \,, \qquad
    \operatorname{Tr}[AB] = \iint A(r,r')\,B(r,r')\,d r\,d r' \,,
\end{equation}
for integral operators $A$ and $B$.

The RPA correlation potential operator depends on both the orbitals and the eigenvalues,
making the electronic ground calculation particularly challenging.
To address this, the OEP formalism~\cite{oepjdtalman, oepkummelperdew,
oepgorlinglevy, engel_dreizler_2011} can be employed, in which the nonlocal
potential operator is replaced by a local, multiplicative potential, as
determined by the variational problem~\cite{oepjdtalman, engel_dreizler_2011}:
\begin{align}
    \min_{V_{s}^{\rm OEP}} \; &E[\widetilde{R}_{nl}, \lambda_{nl}] \quad
    \mathrm{s.t.} \quad
    \Bigg[
        -\frac{1}{2}\frac{d^2}{dr^2}
        +\frac{l(l+1)}{2r^2}
        + V_{s}^{\rm OEP}
    \Bigg]\widetilde{R}_{nl} = \lambda_{nl}\widetilde{R}_{nl} \,,
    \label{eq:EOEPmin}
\end{align}
where $V_s^{\rm OEP} = V_{nuc} + \widetilde{V}_{H}/r + V_{xc}^{\rm OEP}$.
The solution of Eqn.~\ref{eq:EOEPmin} yields the OEP equation for the local
multiplicative potential $V_{xc}^{\rm OEP}$:
\begin{equation}\label{eq:oep-equation}
    \int \widetilde{\chi}_{0,0}(r,r';0)\, V_{xc}^{\rm OEP}(r') \,dr'
    = \Lambda_{x}^{\rm HF}(r) + \Lambda_{c}^{\mathrm{RPA}}(r) \,.
\end{equation}
Here, $\widetilde{\chi}_{0,0}(r,r';0)$ is the static radial density response
function, which along with the right-hand-side terms $\Lambda_{x}^{\mathrm{HF}}$ and
$\Lambda_{c}^{\mathrm{RPA}}$ are given by \cite{trivedi2026spectral}:
\begin{subequations}
\begin{align}
    \widetilde{\chi}_{0,0}(r,r';0)
    &= -2\sum_{nl} g_{nl}\,\widetilde{R}_{nl}(r)
       \sum_{n'\neq n}
       \frac{\widetilde{R}_{n'l}(r)\,\widetilde{R}_{n'l}(r')}
            {\lambda_{n'l}-\lambda_{nl}}
       \,\widetilde{R}_{nl}(r') \,,
    \label{eq:chi0static} \\
    \Lambda_{x}^{\rm HF}(r)
    &= \sum_{nl}\left[
        2\,g_{nl}
        \int
            \frac{\delta \widetilde{R}_{nl}(r')}{\delta V_{s}(r)}\,
            \hat{V}_x^{\rm HF}\widetilde{R}_{nl}(r')\,
        dr'
    \right] \,,
    \label{eq:oep-driving-term-exact-exchange} \\
    \Lambda_{c}^{\mathrm{RPA}}(r)
    &= \sum_{nl}\left[
        \int
            \frac{\delta \widetilde{R}_{nl}(r')}{\delta V_{s}(r)}\,
            \hat{V}_{c}^{\mathrm{RPA}}\widetilde{R}_{nl}(r')\,
        dr'
        + \frac{\delta E_{c}^{\mathrm{RPA}}}{\delta \lambda_{nl}}
          \frac{\delta \lambda_{nl}}{\delta V_{s}(r)}
    \right] \,,
    \label{eq:oep-driving-term-general}
\end{align}
\end{subequations}
where
\begin{equation}\label{eq:oep-second-term-simplified}
    \begin{aligned}
        \hat{V}_{c}^{\mathrm{RPA}}\widetilde{R}_{n l}(r')
        & = -\frac{4}{\pi} \sum_{n' l'} \left(f_{n l}-f_{n' l'}\right)\sum_{l'\!'} (2l'\!'+1) C_{l'\!';l l'}\widetilde{R}_{n' l'}(r') \\
        & \quad \times \int\!\left(D_{n l,\,n' l'}(i\omega)\, \int \widetilde{R}_{n' l'}(r'\!')\,\widetilde{R}_{n l}(r'\!')\,W_{l'\!'}(r'\!',r';\iu\omega)\,dr'\!'\right)d\omega \,,
    \end{aligned}
\end{equation}
with the correlation part of the screened Coulomb interaction given by
\begin{equation}\label{eq:w-screened}
    W_{l'\!'}(\iu\omega) = \nu_{l'\!'}\left[\left(I-\widetilde{\chi}_{0,l'\!'}(\iu\omega)\,\nu_{l'\!'}\right)^{-1}-I\right] \,.
\end{equation}
In addition,
\begin{subequations}
\begin{align}
    \frac{\delta \widetilde{R}_{nl}(r')}{\delta V_s(r)}
    &= -\sum_{n' \neq n}
       \frac{\widetilde{R}_{n'l}(r)\,\widetilde{R}_{n'l}(r')}
            {\lambda_{n'l}-\lambda_{nl}}
       \,\widetilde{R}_{nl}(r) \,,
    \label{eq:dR-dVs-expanded} \\
   \frac{\delta E_{c}^{\mathrm{RPA}}}{\delta \lambda_{n l}}
    &= \frac{2}{\pi} \sum_{n' l'} \left(f_{n l}-f_{n' l'}\right)
    \sum_{l'\!'} (2l'\!'+1) C_{l'\!';l l'}
    \nonumber \\
    &\quad \times \!\int\!\Bigg(D^{(2)}_{n l,\, n' l'}(i\omega) \!\iint
    \widetilde{R}_{n l}(r')\,\widetilde{R}_{n' l'}(r')\,\widetilde{R}_{n l}(r'\!')\,\widetilde{R}_{n' l'}(r'\!')
    W_{l'\!'}(r'\!',r';\iu\omega)\,dr'\,dr'\!'\Bigg)d\omega \,, \label{eq:dExc-dlambda-expanded} \\
    \frac{\delta \lambda_{nl}}{\delta V_s(r)}
    &= \widetilde{R}_{nl}^{2}(r) \,,
    \label{eq:dlambda-dVs}
\end{align}
\end{subequations}
where
\begin{align}
    D^{(2)}_{nl,\,n'l'}(i\omega)
    &= \frac{(\lambda_{nl}-\lambda_{n'l'})^{2}-\omega^{2}}
            {\left[(\lambda_{nl}-\lambda_{n'l'})^{2}+\omega^{2}\right]^{2}} \,.
    \label{eq:D2}
\end{align}
Here, the OEP method has been applied to both the exchange and
correlation components of RPA. The OEP formalism is not limited to RPA --- it can
equally be applied to Hartree--Fock and hybrid calculations, or to
any functional whose potential operator is nonlocal. In the limiting case of a
local potential operator, as in LDA or GGA, the OEP formalism reduces to the
standard Kohn--Sham equations described previously.


\paragraph{Pseudopotential approximation} 
In the pseudopotential approximation, the core electrons are eliminated and the
singular Coulomb potential is replaced by an effective potential.
Consider an atom with $N_v$ valence electrons and a norm-conserving
pseudopotential expressed in Kleinman--Bylander form~\cite{kleinman1980relativistic}.
The energy functional now takes the form~\cite{bhowmik2025spectral}:
\begin{equation} \label{eq:app-energy-psp}
    E[\widetilde{R}_{nl},\lambda_{nl}] = T_s[\widetilde{R}_{nl}]
    + E_{xc}[\rho, \nabla\rho, \tau, \widetilde{R}_{nl},\lambda_{nl}]
    + E_{H}[\rho] + E_{loc}[\rho] + K[\widetilde{R}_{nl}] \,,
\end{equation}
where $\rho$ is the valence electron density, $E_{loc}$ is the ion--electron
interaction energy, and $K$ is the nonlocal pseudopotential energy:
\begin{subequations}
\begin{align}
    E_{loc}[\rho] &= 4\pi \int \rho(r) \, V_{loc}(r) \, r^2 \, dr \,,
    \label{eq:app-Eloc} \\
    K[\widetilde{R}_{nl}] &= \sum_{nl} g_{nl} \sum_{p} \gamma_{lp}
    \left( \int \widetilde{\chi}_{lp}(r)\, \widetilde{R}_{nl}(r) \, dr \right)^2 \,.
    \label{eq:app-K}
\end{align}
\end{subequations}
Here, $V_{loc}$ denotes the local ionic potential, $\chi_{lp} = \widetilde{\chi}_{lp}/r$
is the radial component of the nonlocal projectors, $\gamma_{lp}$ are the
corresponding normalization constants, and $p$ indexes the projectors for
each angular momentum channel $l$. The Euler--Lagrange equations take the
form~\cite{bhowmik2025spectral}:
\begin{subequations}\label{eq:app-el}
\begin{align}
    \left[ -\frac{1}{2}\frac{d^2}{dr^2} + \frac{l(l+1)}{2r^2}
    + V_{loc} + \frac{\widetilde{V}_{H}}{r} + V_{xc}
    + \hat{V}_{NL} \right] \widetilde{R}_{nl}
    &= \lambda_{nl}\widetilde{R}_{nl} \,, \nonumber \\
    \widetilde{R}_{nl}(0) = 0 \,, \quad
    \widetilde{R}_{nl}(r \rightarrow \infty) &= 0 \,;
    \label{eq:app-eigen} \\
    \frac{d^2 \widetilde{V}_{H}(r)}{dr^2} = -4\pi r \rho(r) \,, \quad
    \widetilde{V}_{H}(0) = 0 \,, \quad
    \widetilde{V}_{H}(r \rightarrow \infty) &= N_v \,,
    \label{eq:app-poisson-psp}
\end{align}
\end{subequations}
where the nonlocal pseudopotential operator:
\begin{equation}\label{eq:app-Vnl}
    \hat{V}_{NL}\widetilde{R}_{nl}(r)
    = \sum_{p} \gamma_{lp}\,\widetilde{\chi}_{lp}(r)
      \int \widetilde{\chi}_{lp}(r')\, \widetilde{R}_{nl}(r') \, dr' \,.
\end{equation}
All other components of the energy functional and the Euler--Lagrange equations
remain identical to the all-electron case.

\section{Spectral finite element framework}
\label{Sec:FEM}

\begin{figure}[h!]
    \centering
    \includegraphics[width=0.9\textwidth]{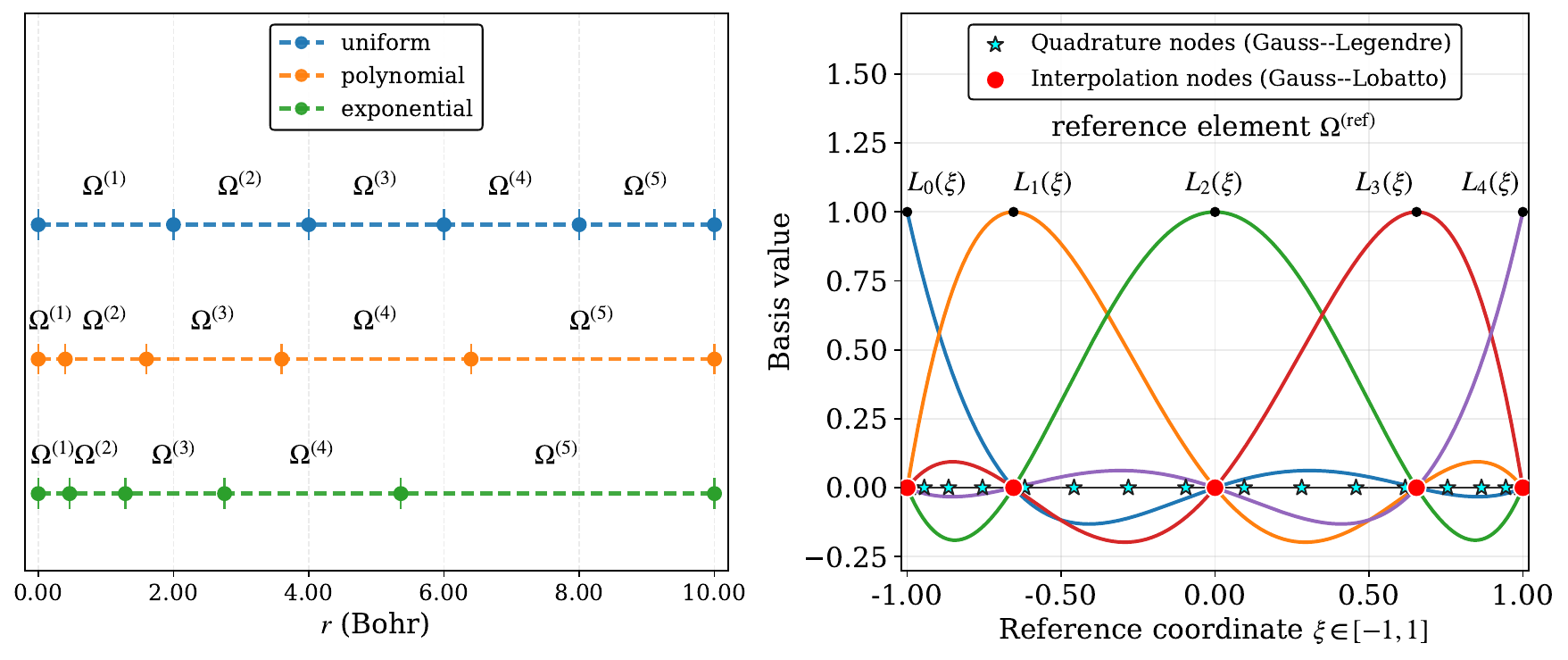}
    \caption{Illustration of the spectral finite-element framework employed in SPARC-atomSFE. Left: domain of $\Omega = [0, R_{\max}]$ partitioned into uniform, polynomial, and exponential meshes with $R_{\max} = 10$~Bohr and $N_{fe} = 5$ elements. Right: reference element $\Omega^{(ref)} = [-1, 1]$, the Legendre--Gauss--Lobatto interpolation nodes of degree $p = 4$, the corresponding Lagrange basis functions, and the Gauss--Legendre quadrature nodes of order $q = 16$.}
    \label{Fig:FEM}
\end{figure}

We now describe the spectral finite-element framework employed in SPARC-atomSFE. The finite-element method is chosen for its systematic improvability, support for high-order approximations, and flexibility in accommodating adaptive grids, properties that together yield an accurate and accurate scheme in the present context. Here, the term \emph{spectral} refers to the use of high-order polynomial approximations in conjunction with appropriately chosen quadrature rules, consistent with its usage in the finite-element literature~\cite{PATERA1984468}. The framework, schematically illustrated in Fig.~\ref{Fig:FEM}, is described in detail next. 

The radial domain is truncated to $\Omega = [0, R_{\max}]$, justified by the exponential decay of the orbitals, and partitioned into non-overlapping subdomains $\{\Omega^{(e)}\}_{e=1}^{N_\text{fe}}$, referred to as finite elements. Within each element $\Omega^{(e)}$, we adopt the basis set expansions:
\begin{subequations}
\begin{align}
    \widetilde{R}_{nl}(r) &= \sum_{j=0}^{p_1} [\mathbf{\widetilde{R}}_{nl}]^{(e)}_{j} L^{p_1(e)}_{j}(r) \,, \label{eq:Rtildediscretization} \\
    \widetilde{V}_{H}(r) &= \sum_{j=0}^{p_2} [\mathbf{\widetilde{V}}_{H}]^{(e)}_{j} L^{p_2(e)}_{j}(r) \,, \label{eq:Vhtildediscretization}
\end{align}
\end{subequations}
where $[\,\cdot\,]^{(e)}_{j}$ denotes the value of the quantity at the $j$-th node of element $e$, and $p_1$ and $p_2$ denote the polynomial degrees of the Lagrange basis functions $L^{p_1(e)}_{j}$ and $L^{p_2(e)}_{j}$, respectively. Each basis function is compactly supported on $\Omega^{(e)}$ and satisfies the Kronecker delta property, so that there are $p_1+1$ and $p_2+1$ nodes per element for $\widetilde{R}_{nl}$ and $\widetilde{V}_{H}$, respectively. The use of different polynomial degrees for $\widetilde{R}_{nl}$ and $\widetilde{V}_{H}$ is motivated by efficiency: $\widetilde{V}_{H}$ has higher frequency content than $\widetilde{R}_{nl}$ and benefits from a higher-degree approximation, while the cost of solving the associated linear system is minor compared to that of the eigenproblem for $\widetilde{R}_{nl}$. Throughout, lowercase indices denote local node numbering within each element and uppercase indices denote global numbering. The mapping from local to global indices for the elements used for $\widetilde{R}_{nl}$ and $\widetilde{V}_{H}$ is denoted by $\mathcal{M}^{p_1}$ and $\mathcal{M}^{p_2}$, respectively, with $\mathcal{M}^{p_1(e)}_{k}$ and $\mathcal{M}^{p_2(e)}_{k}$ providing the global index of the node with local index $k$ in element $e$. Below, we present the discrete weak form of the governing equations presented in the previous section. All integrals so appearing  are evaluated numerically via quadrature of order $q$ on the reference element $\Omega^{(ref)}$, to which each element is mapped.

The discrete form of the angular-momentum-dependent eigenproblem in Eq.~\ref{EL:Eigenproblem} takes the form:
\begin{align}
    \Hm_{l} \! \br_{nl} = \lambda_{nl} \! \M \! \br_{nl}  \,, \quad \Hm_{l} = \hat{\Hm}_{l} + \bvxc \,,  \label{eq:discreteGenEigenProblem}
\end{align}
where the global matrices $\hat{\Hm}_{l}$ and $\M$, and the global vector $\br_{nl}$ are assembled from their element-level counterparts as:
\begin{subequations}
\begin{align}
    [\hat{\Hm}_{l}]_{IJ} &= \sum_{e=1}^{N_{fe}}\sum_{i, j = 0}^{p_1}[\hat{\Hm}_{l}]^{(e)}_{ij} \delta_{I\mathcal{M}^{p_1(e)}_{i}}\delta_{J\mathcal{M}^{p_1(e)}_{j}} \,, \label{eq:assembled_hatHl}\\
    [\M]_{IJ} &= \sum_{e=1}^{N_{fe}}\sum_{i, j = 0}^{p_1}[\M]^{(e)}_{ij} \delta_{I\mathcal{M}^{p_1(e)}_{i}}\delta_{J\mathcal{M}^{p_1(e)}_{j}} \,, \label{eq:assembled_overlap} \\
    [\br_{nl}]_{I} &= \sum_{e=1}^{N_{fe}}\sum_{i = 0}^{p_1}g_i^{(e)}[\br_{nl}]^{(e)}_{i} \delta_{I\mathcal{M}^{p_1(e)}_{i}} \,, \label{eq:assembled_Rnl}
\end{align}
\end{subequations}
with the element-level matrices given by:
\begin{subequations}
\begin{align}
    [\hat{\Hm}_{l}]^{(e)}_{ij} = &  \int \Bigg[\frac{1}{2}\frac{dL^{p_1(e)}_{i}(r)}{dr} \frac{d L^{p_1(e)}_{j}(r)}{d r} + L^{p_1(e)}_{i}(r)\Bigg(\!\frac{l(l+1)}{2r^2}\Bigg)L^{p_1(e)}_{j}(r)  \nonumber \\ 
    & +L^{p_1(e)}_{i}(r)\Bigg(\!V_{nuc}(r) + \!\frac{\widetilde{V}_H(r)}{r}\!\Bigg)L^{p_1(e)}_{j}(r)\Bigg]\, dr \,,
    \label{eq:element_hatHl}\\
    [\M]^{(e)}_{ij}  = &  \int L^{p_1(e)}_{i}(r) L^{p_1(e)}_{j}(r) \, dr \, .  \label{eq:element_overlap}
\end{align}
\end{subequations}
Above, $\delta$ denotes the Kronecker delta, $g_i^{(e)} = 1$ for interior nodes, and $g_i^{(e)} = 0.5$ for nodes shared between adjacent elements. For LDA, GGA, and mGGA, the global matrix $\bvxc$ can be assembled as:
\begin{align}
    [\bvxclgm]_{IJ} &= \sum_{e=1}^{N_{fe}}\sum_{i, j = 0}^{p_1}[\bvxclgm]^{(e)}_{ij} \delta_{I\mathcal{M}^{p_1(e)}_{i}}\delta_{J\mathcal{M}^{p_1(e)}_{j}} \,, \label{eq:assembled_VxcLDAGGAmGGA}
\end{align}
with the element-level matrices given by:
\begin{subequations}
\begin{align}
    [\bvxclgmm]^{(e)}_{ij} =& \int L_i^{p_1(e)}(r) \Big(\hat{V}_{xc}^{\mathrm{LDA/GGA}}\Big)L_j^{p_1(e)}(r) \, dr \,, \label{eq:element_VxcLDAGGAmGGA} \\
    [\bvxcm]^{(e)}_{ij} =& \int L_i^{p_1(e)}(r) \hat{V}_{xc}^{\mathrm{GGA}}L_j^{p_1(e)}(r) \, dr + \int \frac{1}{2}\frac{dL^{p_1(e)}_{i}(r)}{dr}\rho(r)\frac{d\varepsilon_{xc}}{d\tau}(r) \frac{d L^{p_1(e)}_{j}(r)}{d r} \, dr \nonumber \\ &+  \frac{1}{2}\int L_i^{p_1(e)}(r)\Bigg(\frac{1}{r}\frac{d}{dr}\Big[\rho(r)\frac{d\varepsilon_{xc}}{d\tau}(r)\Big]\Bigg) L_{j}^{p_1(e)}\, dr \nonumber \\ &+ \int L_i^{p_1(e)}(r) \Bigg(\rho(r)\frac{d\varepsilon_{xc}}{d\tau}(r)\frac{l(l+1)}{2r^2}\Bigg) L_j^{p_1(e)}(r) \, dr \,, \label{eq:element_VxcmGGA}
\end{align}
\end{subequations}
where the density $\rho$ and its derivatives are evaluated at the quadrature nodes via Eqn.~\ref{density}, using the basis expansion for $\widetilde{R}_{nl}$ given in Eqn.~\ref{eq:Rtildediscretization}. For exact exchange, the global matrix $\bvxhf$ is assembled  as:
\begin{align}
    [\bvxhf]_{IK} &= \sum_{e, e'=1}^{N_{fe}}\sum_{i, k = 0}^{p_1}[\bvxhf]^{(e, e')}_{ik} \delta_{I\mathcal{M}^{p_1(e)}_{i}}\delta_{K\mathcal{M}^{p_1(e')}_{k}} \,, \label{eq:assembled_Vxhf}
\end{align}
with the element-level matrices given by:
\begin{align}
    [\bvxhf]^{(e, e')}_{ik} &= \int L_i^{p_1(e)}(r) \Bigg(-\frac{1}{2} \sum_{n'l'} g_{n'l'} \sum_{l'\!'=|l-l'|}^{l+l'} \left[
    \begin{pmatrix}
            l & l' & l'\!' \\
            0 & 0 & 0
    \end{pmatrix}^2
    \widetilde{R}_{n'l'}(r) \frac{Y_{kn'l'}^{l'\!' (e')}(r)}{r} \right]\Bigg) \, dr \,. \label{eq:element_Vxhf}
\end{align}
Above, $Y_{kn'l'}^{l'\!' (e')}$ is given by the basis set expansion:
\begin{align}
    Y_{kn'l'}^{l'\!' (e')}(r) &= \sum_{j=0}^{p_2}[\bY_{kn'l'}^{l'\!' (e')}]^{(e)}_{j}L_j^{p_2(e)}(r) \, , \label{Ydiscretization}
\end{align}
whose coefficients are determined by solving the linear system:
\begin{align}
    \bD_{l'\!'}\bY_{kn'l'}^{l'\!' (e')} &= \bP_{kn'l'}^{l'\!' (e')} \,,  \label{eq:discreteYdiff}
\end{align}
where the global matrices/vectors are assembled as:
\begin{subequations}
    \begin{align}
        [\bD_{l'\!'}]_{IJ} &= \sum_{e=1}^{N_{fe}}\sum_{i,j=0}^{p_2}[\bD_{l'\!'}]^{(e)}_{ij} \delta_{I\mathcal{M}^{p_2(e)}_{i}}\delta_{J\mathcal{M}^{p_2(e)}_{j}} \,, \label{eq:assembled_D} \\
    [\bP_{kn'l'}^{l'\!' (e')}]_{I} &=   \sum_{e=1}^{N_{fe}}\sum_{i =0}^{p_2}[\bP_{kn'l'}^{l'\!' (e')}]^{(e)}_{i} \delta_{I\mathcal{M}^{p_2(e)}_{i}}  \,,
    \label{eq:assembled_P}
    \\
    [\bY_{kn'l'}^{l'\!' (e')}]_{I} &= \sum_{e=1}^{N_{fe}}\sum_{i = 0}^{p_2}g_i^{(e)}[\bY_{kn'l'}^{l'\!' (e')}]^{(e)}_{i} \delta_{I\mathcal{M}^{p_2(e)}_{i}}
   \,,
    \label{eq:assembled_Y}
    \end{align}
\end{subequations}
with the element-level matrices/vectors given by:
\begin{subequations}
    \begin{align}
        [\bD_{l'\!'}]^{(e)}_{ij} &= - \int \Bigg(\frac{dL_{i}^{p_2(e)}(r)}{dr} \frac{dL_{j}^{p_2(e)}(r)}{d r} + L_{i}^{p_2(e)}(r)\frac{l'\!'(l'\!'+1)}{r^2}L_{j}^{p_2(e)}(r) \Bigg)\,dr\,, \label{eq:element_D}\\
        [\bP_{kn'l'}^{l'\!' (e')}]^{(e)}_{i} &= -(2l'\!'+1)\int L_{i}^{p_2(e)}(r) \frac{\widetilde{R}_{n'l'}(r)}{r} L_k^{p_1(e')}(r) \, dr\, . \label{eq:element_P}
    \end{align}
\end{subequations}
The Dirichlet boundary condition at $r=0$ is enforced by removing the first row and column of matrix $\bD_{l'\!'}$, along with the first entries of vectors $\bY_{kn'l'}^{l'\!' (e')}$ and $\bP_{n'l'}^{l'\!' (e')}$. The Robin boundary condition at $r=R_{max}$ is enforced by adding $-l'\!'/R_{max}$ to the diagonal entry corresponding to the last row and column of $\bD_{l'\!'}$. It is worth noting that we employ the differential-equation form of the exact exchange operator (Eq.~\ref{Vxx2}) rather than its integral counterpart (Eq.~\ref{Vxx}), as the former requires a significantly lower quadrature order. Once the Hamiltonian matrix $\Hm_{l}$ is constructed as described above, homogeneous Dirichlet boundary conditions on $\widetilde{R}_{nl}(r)$ are enforced by removing the first and last rows and columns of both $\Hm_l$ and $\M$, together with the first and last entries of $\br_{nl}$.

The discrete form of the Poisson problem in Eqn.~\ref{EL:Poisson} takes the form:
\begin{align}
    \LL\bvh & = \F \, , \label{eq:FEPossion} 
\end{align}
where the global matrix/vectors are assembled from their element-level counterparts as:
\begin{subequations}
\begin{align}
    [\LL]_{IJ} &= \sum_{e=1}^{N_{fe}}\sum_{i,j=0}^{p_2}[\LL]^{(e)}_{ij} \delta_{I\mathcal{M}^{p_2(e)}_{i}}\delta_{J\mathcal{M}^{p_2(e)}_{j}} \,, \label{eq:assembled_Laplacian} \\
    [\F]_{I} &=   \sum_{e=1}^{N_{fe}}\sum_{i=0}^{p_2}[\F]^{(e)}_{i} \delta_{I\mathcal{M}^{p_2(e)}_{i}} \,, 
    \label{eq:assembled_PoissonRHS} 
    \\
    [\bvh]_{I} &= \sum_{e=1}^{N_{fe}}\sum_{i = 0}^{p_2}g_i^{(e)}[\bvh]^{(e)}_{i} \delta_{I\mathcal{M}^{p_2(e)}_{i}} \,, 
    \label{eq:assembled_Vh}
\end{align}
\end{subequations}
with the element matrices/vectors given by:
\begin{subequations}
\begin{align}
    [\LL]^{(e)}_{ij} &= \int \frac{dL_{i}^{p_2(e)}(r)}{dr} \frac{dL_{j}^{p_2(e)}(r)}{d r} \,dr\,,
    \label{eq:element_Laplacian}\\
    [\F]^{(e)}_{i} &= 4\pi \int r \rho(r) L_{i}^{p_2(e)}(r)  \,dr\,.
    \label{eq:element_PoissonRHS}
\end{align}
\end{subequations}
Here, the density $\rho$ is evaluated at the quadrature nodes via Eqn.~\ref{density}, using the basis expansion for $\widetilde{R}_{nl}$ given in Eqn.~\ref{eq:Rtildediscretization}. The homogeneous Dirichlet boundary condition on $\widetilde{V}_{H}$ at $r=0$ is enforced by removing the first row and column of matrix $\LL$ (Eq.~\ref{eq:assembled_Laplacian}), along with the first entries of vectors $\bvh$ and $\F$. The nonhomogeneous Dirichlet boundary condition at $r=R_{max}$ is imposed by removing the last row and column of $\LL$ (Eq.~\ref{eq:assembled_Laplacian}) and the last entries of $\bvh$ and $\F$, with $\F$ updated accordingly to account for the boundary value, and the last entry of $\bvh$ set to the number of electrons $N_e$.


\paragraph{RPA--OEP}  We adopt the following basis set expansion for the OEP potential:
\begin{align}
    V_{xc}^{\rm OEP}(r) &= \sum_{j=0}^{p_3} [\mathbf{\bvxcoep}]^{(e)}_{j} L^{p_3(e)}_{j}(r) \,, \label{eq:Vxcoepdiscretization}
\end{align}
where the polynomial degree $p_3$ is typically chosen to be smaller than $p_1$, the degree used for $\widetilde{R}_{nl}$,  for numerical stability \cite{scRPAgorlingpaper, trivedi2026spectral}. The discrete form of the linear system for the OEP potential, Eqn.~\ref{eq:oep-equation}, takes the form:
\begin{align}
    \chizero \bvxcoep & = \blambdaxhf + \blambdacrpa\, , \label{eq:discreteOEP}
\end{align}
where the global matrix/vectors are assembled from the element-level counterparts as:
\begin{subequations}
    \begin{align}
        [\chizero]_{IJ} &= \sum_{e, e'=1}^{N_{fe}}\sum_{i, j=0}^{p_3}[\chizero]^{(e, e')}_{ij} \delta_{I\mathcal{M}^{p_3(e)}_{i}}\delta_{J\mathcal{M}^{p_3(e')}_{j}} \,, \label{eq:assembled_chi0}\\
        [\blambdaxhf]_{I} &= \sum_{e=1}^{N_{fe}}\sum_{i=0}^{p_3}[\blambdaxhf]^{(e)}_{i} \delta_{I\mathcal{M}^{p_3(e)}_{i}} \,, \label{eq:assembled_OEPxRHS}\\
        [\blambdacrpa]_{I} &= \sum_{e=1}^{N_{fe}}\sum_{i=0}^{p_3}[\blambdacrpa]^{(e)}_{i} \delta_{I\mathcal{M}^{p_3(e)}_{i}} \,, \label{eq:assembled_OEPcRHS}
    \end{align}
\end{subequations}
with the element matrices/vectors given by:
\begin{subequations}
    \begin{align}
[\chizero]^{(e, e')}_{i, j} =&  \iint L_{i}^{p_3(e)}(r) \chi_{0,0}(r,r';0)L_{j}^{p_3(e')}(r')\, drdr' \, ,\label{eq:element_chi0} \\
[\blambdaxhf]^{(e)}_{i} =& \int L_{i}^{p_3(e)}(r)\Lambda_{x}^{\mathrm{HF}}(r)\, dr \, . \label{eq:element_OEPxRHS}\\
[\blambdacrpa]^{(e)}_{i} =& \int L_{i}^{p_3(e)}(r)\Lambda_{c}^{\mathrm{RPA}}(r)\, dr \, . \label{eq:element_OEPcRPA}
\end{align}
\end{subequations}
Here, the static response function $\widetilde{\chi}_{0,0}$ is calculated on the quadrature nodes using Eqn.~\eqref{eq:chi0static}, with the basis expansion of $\widetilde{R}_{nl}$ given in Eqn.~\eqref{eq:Rtildediscretization}. Similarly, $\Lambda_{x}^{\mathrm{HF}}$ and $\Lambda_{c}^{\mathrm{RPA}}$ are calculated using Eqns.~\eqref{eq:oep-driving-term-exact-exchange} and~\eqref{eq:oep-driving-term-general}, again employing the basis expansion of $\widetilde{R}_{nl}$.

\paragraph{Pseudopotential approximation} In forming the Hamiltonian matrix $\Hm_{l}$, the above formalism is applicable, except that $V_{nuc}$ is replaced by $V_{loc}$. In addition, the global nonlocal pseudopotential matrix must be added to the Hamiltonian, and is assembled from its element-level counterparts as:
\begin{align}
    [{\bvnl}]_{IJ} &= \sum_{e, e'=1}^{N_{fe}}\sum_{i, j=0}^{p_1}[{\bvnl}]^{(e, e')}_{ij} \delta_{I\mathcal{M}^{p_1(e)}_{i}}\delta_{J\mathcal{M}^{p_1(e')}_{j}} \,, \label{eq:assembled_hatHlpsp}
\end{align}
where the element matrices are given by:
\begin{align}
    [{\bvnl}]^{(e, e')}_{ij} =  \int L_{i}^{p_1(e)}(r) \hat{V}_{NL}L_{j}^{p_1(e')}(r)\, dr \,.
    \label{eq:element_hatHlpsp}
\end{align}
Also, in solving the Poisson equation for the Hartree potential, the boundary condition must be updated from $N_e$ to $N_v$.

\paragraph{Energy} Once the electronic ground state has been determined, the total energy can be evaluated using numerical quadrature and the basis set expansions for the various quantities presented above.


\section{Implementation} \label{Sec:Implementation}

The SPARC-atomSFE code is implemented in Python, making it  well-suited to rapid 
prototyping and seamless integration with modern machine learning frameworks 
and workflows. Three mesh types are supported for 
the placement of finite-element nodes: uniform, polynomial~\cite{gwasphericalatomshellgren}, 
and exponential~\cite{vcertik2024high, lehtola2019fully}:
\begin{subequations}
\begin{align}
    \text{uniform:}\quad
    r_i &= \frac{i}{N_{fe}}R_{max} \,, \\
    \text{polynomial:}\quad
    r_i &= R_{\max}\left(\frac{i}{N_{fe}}\right)^s \,, \qquad s>0 \,, \\
    \text{exponential:}\quad
    r_i &= c + \alpha\!\left(e^{\beta i}-1\right) \,, \quad
    \beta=\frac{\ln a}{N_{fe}-1} \,, \quad
    \alpha=\frac{R_{max}-c}{e^{\beta N_{fe}}-1} \,, \label{eq:mesh-exponential}
\end{align}
\end{subequations}
where $i=0,\ldots,N_{fe}$. For polynomial meshes, $s>1$ clusters nodes near the origin; 
for exponential meshes, $c\geq 0$ is the shift and $a>0$ is the concentration parameter. 
Independent polynomial degrees can be chosen for the orbitals ($p_1$) and the OEP 
potential ($p_3$); the polynomial degree for the Hartree potential is set to 
$p_2 = 2p_1 + 1$, and typically $p_3 \approx p_1/4$ is chosen.  Arbitrary polynomial degrees are supported, though in practice $p_1 \geq 40$ 
tends to lead to numerical instability. Spatial integrals are evaluated using Gauss-Legendre quadrature. In  RPA-OEP, $\omega$-frequency integration is performed using Gauss-Legendre 
quadrature, whose order is denoted by $N_{\omega}$, and the summation over angular momentum is truncated at a maximum 
value $l_{max}$.

SPARC-atomSFE supports both all-electron and pseudopotential calculations, with the latter employing Optimized Norm-Conserving Vanderbilt (ONCV) pseudopotentials~\cite{hamann2013optimized} in the \texttt{.psp8} format, optionally including nonlinear core corrections. The package provides a broad range of exchange-correlation approximations, including LDA with Slater exchange and Vosko--Wilk--Nusair~\cite{LDA_VWN}, Perdew--Zunger~\cite{perdew1981self}, or Perdew--Wang~\cite{perdew1992accurate} correlation; GGA with the Perdew--Burke--Ernzerhof (PBE) functional~\cite{perdew1996generalized}; meta-GGA with the Strongly Constrained and Appropriately Normed (SCAN)~\cite{sun2015strongly}, rSCAN~\cite{bartok2019regularized}, and r$^2$SCAN~\cite{r2SCAN} functionals; hybrid functionals, including PBE0~\cite{adamo1999toward} and a generalized PBE0 form with variable exact-exchange mixing; Hartree--Fock (HF) theory~\cite{Fock1930126};  and RPA-OEP. Charged atoms and fractional occupations are supported across all these cases.

The electronic ground state is obtained using the self-consistent field (SCF) method.
For functionals involving nonlocal exchange-correlation potentials, such as HF and PBE0, an outer--inner loop structure is employed: the outer loop performs a fixed-point iteration with respect to the nonlocal potential operator, which is held fixed during each inner loop, while the inner loop performs a fixed-point iteration with respect to the electron density.
For RPA-OEP calculations, the same outer--inner loop strategy is used, with the outer loop performing a fixed-point iteration with respect to the OEP exchange-correlation potential. The initial guess for the orbitals in the outer loop is taken from a fully converged 
PBE calculation or from the first PBE iteration, while the initial electron density 
is obtained from the Thomas-Fermi approximation or from PBE when available. Inner-loop convergence is accelerated using the the periodic Pulay mixing scheme~\cite{banerjee2016periodic}, with the option of the RPA dielectric matrix as a preconditioner. Direct linear solvers are employed for the Poisson, exact exchange, and OEP equations.
Direct eigensolvers are used for all functionals except SCAN and r$^2$SCAN, for which an iterative eigensolver is employed through Hamiltonian-vector products, as this was found to be more stable.

The OEP equation for the local exchange-correlation potential 
is solved up to an indeterminate constant. The RPA-OEP potential is therefore 
shifted by a constant to match $-1/r$ at a specified distance \cite{vacondiopaper}, e.g., $9$~bohr in 
practice, and the potential beyond this distance is set to 
$-1/r$, consistent with the asymptotic decay of the exact exchange potential that 
dominates at large $r$. This correction is applied 
to avoid numerical instabilities that can manifest as oscillations near the boundary 
for large domain sizes, a consequence of the singular nature of the density response matrix. The quantities $\Lambda_c^{\mathrm{RPA}}$, $E_c^{\mathrm{RPA}}$, 
and the corresponding energy density are computed via a thread-based embarrassingly 
parallel implementation over the frequency quadrature points, using Python's 
built-in \texttt{concurrent.futures} module. To prevent thread oversubscription,  
 the \texttt{threadpoolctl} library can be used, which restricts NumPy's 
BLAS calls to a single thread during the parallel section.


\section{Results and discussion} \label{Sec:Results}
We now verify the accuracy and performance of SPARC-atomSFE for atomic structure calculations, starting from the 
radial Schrödinger equation and then proceeding to all-electron and pseudopotential Kohn-Sham DFT calculations, 
covering the full range of exchange-correlation functionals: local, semilocal, and 
nonlocal. The data corresponding to the results presented here is available in the SPARC-atomSFE repository.

\subsection{Radial Schr\"odinger}
\begin{figure}[h!]
    \centering
    \includegraphics[width=0.4\textwidth]{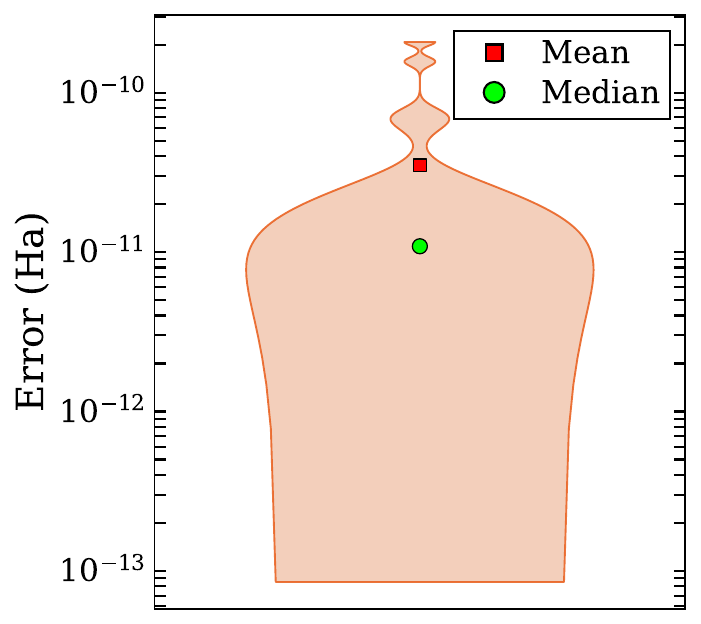}
    \caption{Distribution of eigenvalue errors for the radial Schr\"odinger equation corresponding to $Z=92$.}
    \label{fig:Schrodinger}
\end{figure}

We first verify the accuracy of the spectral finite-element framework for the all-electron radial Schr\"odinger equation~\cite{vcertik2024high, vcertik2013dftatom}, which is obtained by setting $\hat{V}_{xc}=\widetilde{V}_H=0$ in the radial Kohn--Sham equation, Eq.~\eqref{EL:Eigenproblem}, and admits analytical solutions. We consider $Z=92$ and compute all occupied eigenvalues using a domain size of $R_{\max}=40$~Bohr, an exponential mesh with shift $c=0$ and concentration parameter $a=100$, $N_{fe}=12$ finite elements, polynomial degree $p=20$, and quadrature order $q=60$. Fig.~\ref{fig:Schrodinger} shows the eigenvalue errors relative to the analytical values, demonstrating accuracies better than $10^{-10}$~Ha and confirming the accuracy of the spectral finite-element framework.

\subsection{All electron Kohn-Sham DFT}
We next verify the accuracy and performance of SPARC-atomSFE for all-electron Kohn-Sham DFT calculations. Unless otherwise specified, we use an exponential mesh with shift $c=0$, concentration parameter $a=100$, domain size $R_{max}=40$~Bohr, $N_{fe}=12$ finite elements, polynomial degree $p=20$, and quadrature order $q=60$.

\begin{figure}[h!]
    \centering
    \includegraphics[width=0.9\textwidth]{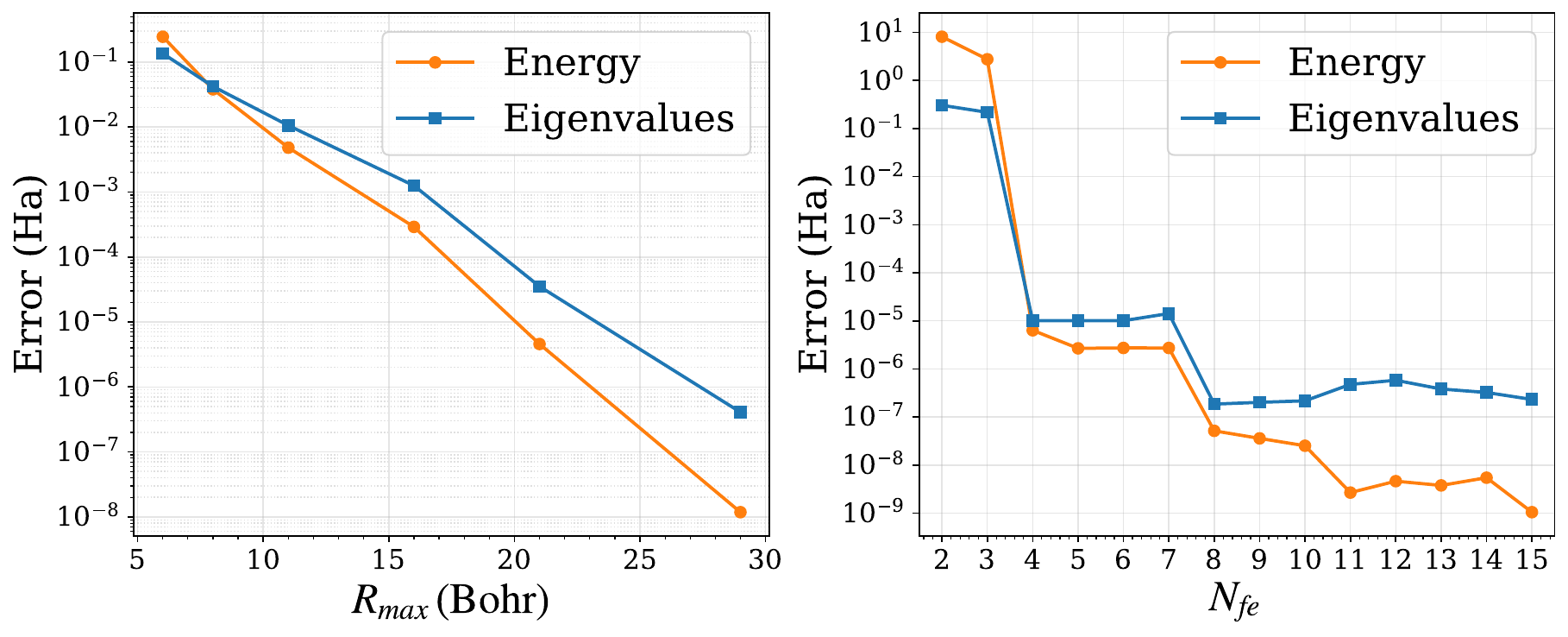}
    \caption{Convergence of the total energy and occupied eigenvalues with respect to domain size 
$R_{max}$ (left) and number of finite elements $N_{fe}$ (right) in all-electron Kohn-Sham DFT calculations, using the PBE exchange-correlation functional, for atomic numbers $Z=1$--$92$. The energy error is defined as the maximum absolute total-energy difference over all atoms, while the eigenvalue error is defined as the maximum, over all atoms, of the per-atom mean absolute difference in the occupied eigenvalues. Reference values are computed using domain size $R_{\max}=40$~Bohr and $N_{fe}=16$ finite-elements.}
    \label{fig:All_electron_convergence}
\end{figure}

We begin by studying convergence with respect to the domain size $R_{max}$ and the number of finite elements $N_{fe}$. In particular, Fig.~\ref{fig:All_electron_convergence} shows the convergence of the total energy and occupied eigenvalues for the PBE functional across atomic numbers $Z=1$--$92$, with reference results computed using $R_{max}=40$~Bohr and $N_{fe}=16$. We observe that both quantities converge exponentially with increasing $R_{max}$, with errors below $10^{-6}$~Ha achieved at $R_{max}\sim 27$~Bohr. We also observe that there is rapid convergence with increasing $N_{fe}$, with accuracies better than $10^{-6}$~Ha attained at $N_{fe}\sim 8$. These results are consistent with previous LDA results obtained using a high-order finite-element framework~\cite{vcertik2024high}; however, GGA requires a somewhat larger number of degrees of freedom, with LDA achieving approximately an order of magnitude higher accuracy for comparable discretization parameters. We observe similar convergence behavior for all other exchange--correlation functionals considered here, except RPA-OEP, which exhibits slower convergence with respect to the discretization parameters. This is expected, given its dependence on both occupied and unoccupied states.

\begin{figure}[h!]
    \centering
    \includegraphics[width=0.75\textwidth]{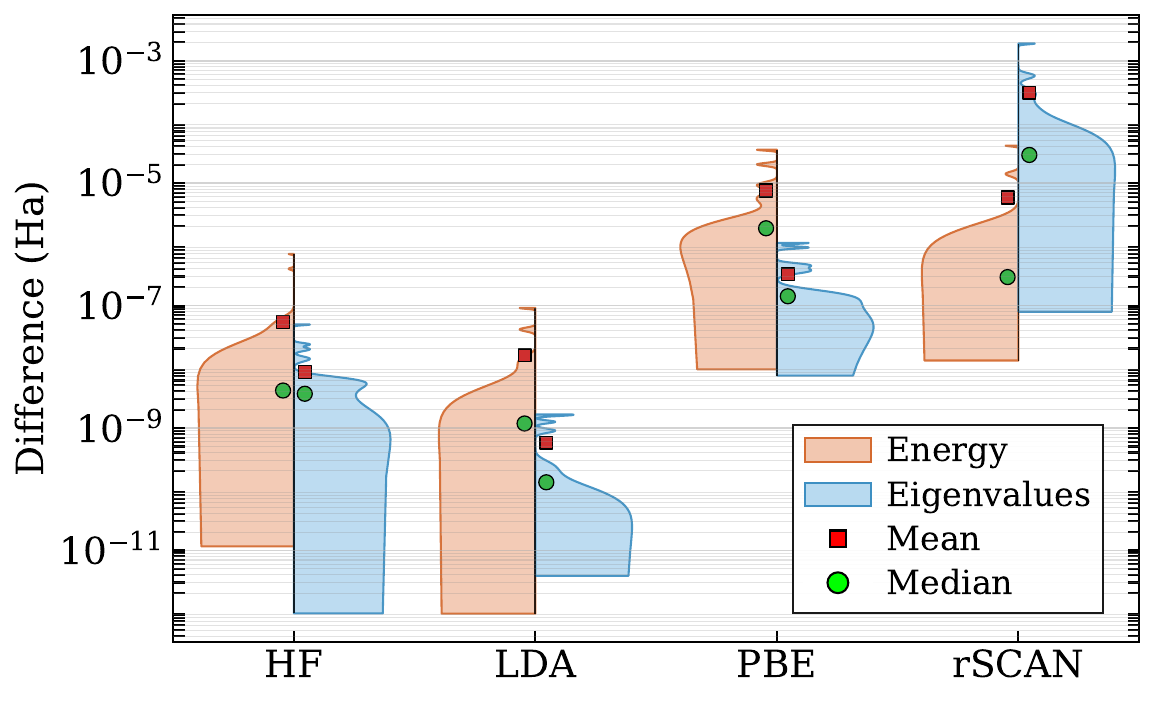}
    \caption{Comparison of SPARC-atomSFE all-electron Kohn-Sham DFT results with literature for various exchange-correlation functionals. The energy and eigenvalue errors denote the absolute differences in the corresponding quantities.}
    \label{fig:All_electron_accuracy}
\end{figure}

We next assess the accuracy of SPARC-atomSFE for all-electron Kohn-Sham DFT calculations through 
comparisons with literature. In particular, we consider four exchange-correlation functionals: LDA, 
PBE, rSCAN, and HF. For LDA, PBE, and rSCAN, results are compared for ten 
atoms ranging from light to heavy --- H, Be, C, Ne, Na, Si, Fe, Kr, Gd, and U --- 
against the featom code~\cite{vcertik2024high} for LDA and the atomPAW 
code~\cite{holzwarth2022cubic} for PBE and rSCAN. For HF, comparisons are 
carried out for closed-shell neutral atoms against Ref.~\cite{cinal2020highly}, and 
for charged species --- anions H$^{-}$, Li$^{-}$, F$^{-}$, Na$^{-}$, Cl$^{-}$ and 
cations Li$^{+}$, B$^{+}$, Na$^{+}$, Al$^{+}$ --- against Ref.~\cite{lehtola2019fully}. 
 The distribution of the differences in the results is shown as violin plots in 
Fig.~\ref{fig:All_electron_accuracy}. We observe that total energies and occupied eigenvalues agree 
to $10^{-6}$~Ha or better in most cases; the somewhat larger errors observed for PBE 
and rSCAN can likely be attributed to the accuracy of the reference results (as we were unable to converge them further) rather than 
to SPARC-atomSFE itself, with rSCAN showing slightly worse agreement than PBE. 
Notably, for LDA, the total energy and eigenvalues of SPARC-atomSFE and featom agree 
to $10^{-8}$~Ha and $10^{-9}$~Ha, respectively, even for the heaviest atom considered, 
U ($Z=92$). These results demonstrate the accuracy of SPARC-atomSFE for  
atomic structure calculations based on all-electron Kohn-Sham DFT.

We next assess the accuracy of SPARC-atomSFE for all-electron RPA-OEP calculations, considering 
the atoms He, Be, and Ne, the reference values for which are available in the 
literature~\cite{trivedi2026spectral, gwasphericalatomshellgren}. We employ a polynomial 
mesh with $s=2$, $R_{max}=13$~bohr, $p_1=15$, $p_3=4$, and $q=55$; $N_{fe}$, 
$l_{max}$, and $N_{\omega}$ are $(20, 18, 30)$, $(30, 18, 40)$, and $(30, 40, 60)$ 
for He, Be, and Ne, respectively, which converges energies and eigenvalues to within $5 \times 10^{-4}$ Ha. As shown in Table~\ref{Table:RPA}, the results 
agree with the literature to within $10^{-3}$~Ha, which is the accuracy typically 
sought/attained in such calculations. Differences with Ref.~\cite{trivedi2026spectral} are a 
consequence of the asymptotic correction applied here, whereby the OEP potential is 
replaced by $-1/r$ beyond $9$~bohr to ensure correct asymptotic decay and numerical 
stability; in these cases, our results agree more closely with the cubic-spline 
reference of Ref.~\cite{gwasphericalatomshellgren} and the complete basis-set extrapolated 
results of Ref.~\cite{scRPAgorlingpaper}.

\begin{table}[htbp]
\centering
\begin{tabular}{ccccc}
\toprule
& & He & Be & Ne \\
\midrule
\multirow{3}{*}{Ionization potential (Ha)}
& SPARC-atomSFE & -0.902 & -0.356 & -0.796 \\
& FE & -0.902 & -0.356 & -0.797 \\
& CS & -0.902 & -0.354 & -0.796 \\
\midrule
\multirow{3}{*}{Energy (Ha)}
& SPARC-atomSFE & -2.945 & -14.754 & -129.146 \\
& FE & -2.945 & -14.754 & -129.147 \\
& CS & -2.945 & -14.754 & -129.143 \\
\bottomrule
\end{tabular}
\caption{Comparison of RPA-OEP ionization potentials and total energies obtained with SPARC-atomSFE against values in the literature. The finite-element results used in prior work~\cite{trivedi2026spectral} are labeled FE, while the cubic-spline results reported in Ref.~\cite{gwasphericalatomshellgren} are labeled CS.}
\label{Table:RPA}
\end{table}

Finally, we assess the performance of SPARC-atomSFE for all-electron Kohn-Sham DFT 
calculations. Table~\ref{tab:timing-accuracy-ae} reports the CPU time  for Au ($Z=79$) across various exchange-correlation functionals to achieve $\sim 10^{-6}$ Ha accuracy, with all 
calculations performed on an Acer Nitro AN515-46 laptop equipped with an AMD Ryzen 7 
6800H CPU and 32~GB RAM, running Windows 11 Home (64-bit). Local and semilocal 
functionals, i.e., LDA, PBE, and rSCAN, complete in well under a second. PBE0 and HF, which require an outer-inner SCF loop structure, 
take on the order of a few seconds due to the additional cost of exact exchange and 
the outer loop iterations. The LDA timing of $0.12$~s ($n_{in}=13$) is comparable to the $0.13$~s ($n_{in}=18$) taken by 
featom~\cite{vcertik2024high} under identical discretization parameters and hardware, 
highlighting the computational efficiency of SPARC-atomSFE despite being implemented 
in Python rather than Fortran. It is worth noting that RPA-OEP calculations are significantly more expensive than lower-rung functionals, requiring more than $1000$~s of CPU time even for the He atom.

\begin{table}[h!]
    \centering
    \begin{tabular}{@{}lccccc@{}}
        \toprule
         & $N_{fe}$ & $n_{out}$ & $n_{in}$ & Time (s) & Error (Ha) \\
        \midrule
        LDA   & $4$ & \textemdash & $13$ & $0.12$ & $3\times 10^{-6}$ \\
        PBE   & $4$ & \textemdash & $11$ & $0.11$ & $5\times 10^{-6}$ \\
        rSCAN & $9$ & \textemdash & $14$ & $0.88$ & $8\times 10^{-6}$ \\
        PBE0  & $4$ & $6$         & $36$ & $4.12$ & $3\times 10^{-6}$ \\
        HF    & $4$ & $11$        & $143$ & $7.76$ & $1\times 10^{-6}$ \\
        \bottomrule
    \end{tabular}
    \caption{CPU time for all-electron Kohn-Sham DFT calculations of 
    Au ($Z=79$) for various exchange-correlation functionals, where $N_{fe}$ is chosen 
    to yield the listed accuracy. $n_{out}$ and $n_{in}$ denote the number of outer 
    and the total number of inner SCF iterations summed over all outer iterations, respectively.}
    \label{tab:timing-accuracy-ae}
\end{table}


\subsection{Pseudopotential Kohn-Sham DFT}

We now verify the accuracy and performance of SPARC-atomSFE for pseudopotential Kohn-Sham DFT 
calculations. Unless otherwise specified, we use an exponential mesh with shift $c=0$, concentration 
parameter $a=20$, domain size $R_{max}=40$~Bohr, $N_{fe}=10$ finite elements, 
polynomial degree $p=20$, and quadrature order $q=60$. We employ two suites of ONCV pseudopotentials: the SPMS set~\cite{shojaei2023soft, hamann2013optimized}, which includes nonlinear core corrections and is used for the LDA, PBE, and PBE0 calculations, and the SG15 set \cite{schlipf2015optimization}, which does not include nonlinear core corrections and is used for the rSCAN calculations. 

\begin{figure}[h!]
    \centering
    \includegraphics[width=0.9\textwidth]{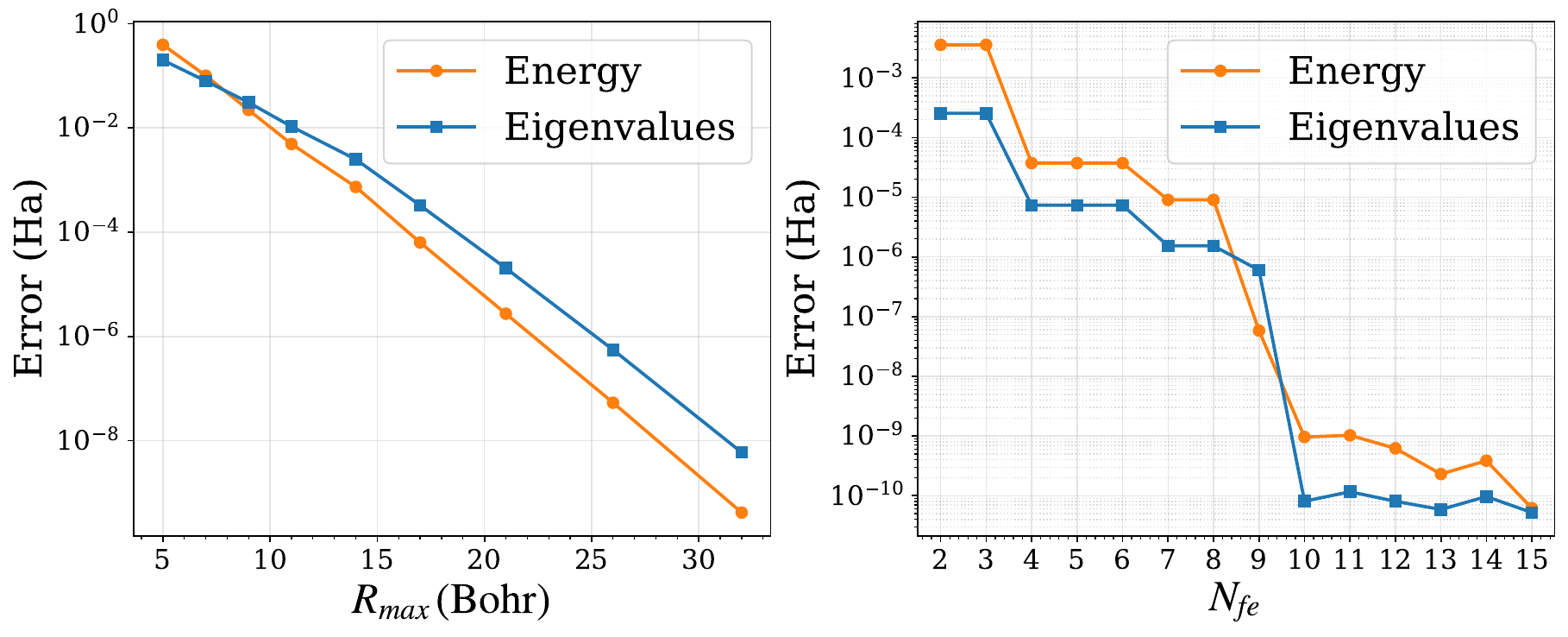}
    \caption{Convergence of the total energy and occupied eigenvalues with respect to domain size 
$R_{max}$ (left) and number of finite elements $N_{fe}$ (right) in pseudopotential Kohn-Sham DFT calculations, using the PBE exchange-correlation functional, for atomic numbers $Z=1$ to $57$ and $Z=72$ to $83$. The energy error is defined as the maximum absolute total-energy difference over all atoms, while the eigenvalue error is defined as the maximum, over all atoms, of the per-atom mean absolute difference in the occupied eigenvalues. Reference values are computed using $R_{\max}=40$~Bohr and $N_{fe}=16$.}
    \label{fig:Pseudopotential_convergence}
\end{figure}

We begin by studying convergence with respect to domain size $R_{max}$ and 
number of finite elements $N_{fe}$.  In particular, Fig.~\ref{fig:Pseudopotential_convergence} shows the convergence of total 
energies and occupied valence eigenvalues for the PBE functional across atomic numbers 
$Z=1$--$57$ and $Z=72$--$83$, with reference results corresponding to 
$R_{max}=40$~Bohr and $N_{fe}=16$. We observe that both quantities converge 
exponentially with increasing $R_{max}$, with errors below $10^{-6}$~Ha achieved again at 
$R_{max} \sim 27$~Bohr. We also observe that there is rapid convergence with increasing 
$N_{fe}$, with accuracies better than $10^{-6}$~Ha attained at 
$N_{fe} \sim 9$. These results are commensurate with previous GGA results 
obtained using a spectral scheme based on Chebyshev 
polynomials~\cite{bhowmik2025spectral}. We also find similar convergence behavior for all other exchange--correlation functionals considered here, with the exception of RPA-OEP, which has not been tested in the pseudopotential context.

\begin{figure}[h!]
    \centering
    \includegraphics[width=0.75\textwidth]{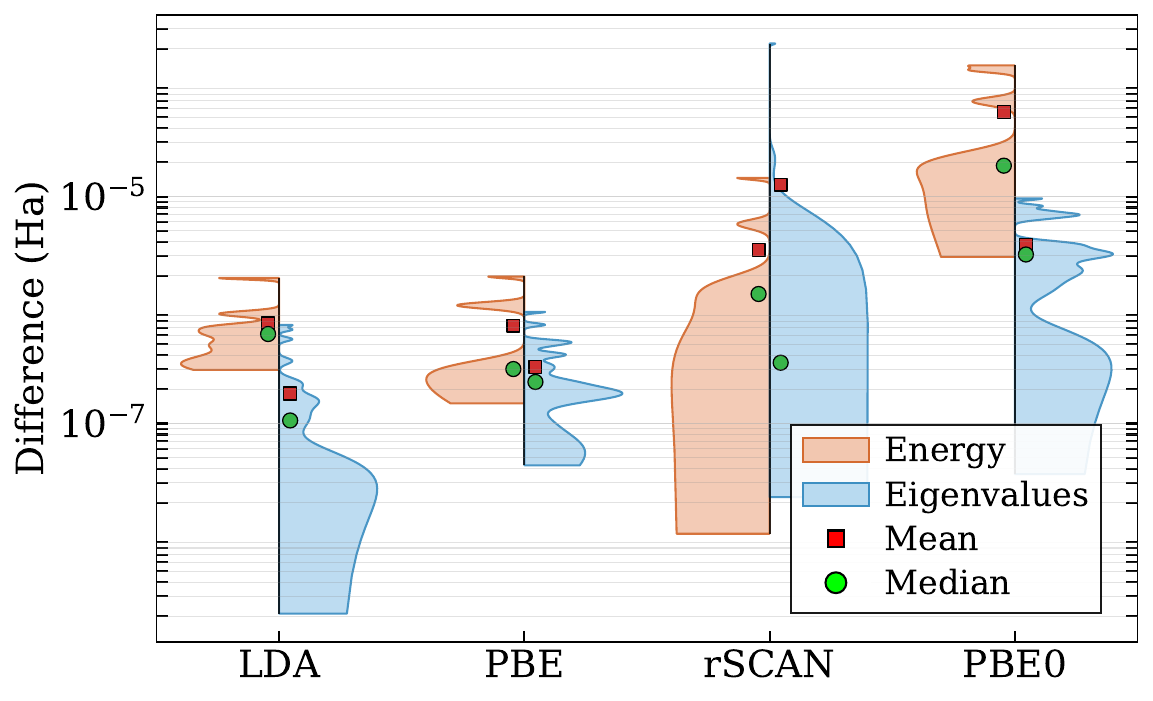}
    \caption{Comparison of SPARC-atomSFE all-electron Kohn-Sham DFT results with literature for various exchange-correlation functionals. The energy and eigenvalue errors denote the absolute differences in the corresponding quantities.}
    \label{fig:Pseudopotential_accuracy}
\end{figure}

We next assess the accuracy of SPARC-atomSFE for pseudopotential Kohn-Sham DFT calculations 
through comparisons with the literature. In particular, we compare against the spectral scheme based SPARC-atom code~\cite{bhowmik2025spectral},  implemented as part of the M-SPARC package~\cite{xu2020m}, while considering four exchange-correlation functionals: LDA, PBE, rSCAN, and PBE0. We compare the results  for seven atoms spanning a range of chemical environments --- He, N, O, 
Fe, Mn, Mo, and Cs.  The distribution of the differences in the results is shown as violin plots in Fig.~\ref{fig:Pseudopotential_accuracy}. We observe that total energies and 
occupied valence eigenvalues agree to $10^{-6}$~Ha or better for LDA, PBE, and rSCAN. 
The exception is PBE0, which exhibits errors approximately an order of magnitude 
larger; this  can be attributed to the fact that the quadrature in the spectral scheme used by SPARC-atom
is limited to the degree of the polynomial used, making it difficult to converge 
calculations involving exact exchange, where higher-order quadrature is required. 
This limitation can be overcome by adopting a differential 
approach for the exact exchange operator, as done in the present work. These results demonstrate 
the accuracy of SPARC-atomSFE for atomic structure calculations based on pseudopotential Kohn-Sham DFT.

Finally, we assess the performance of SPARC-atomSFE for pseudopotential Kohn-Sham 
DFT calculations. Table~\ref{tab:timing-accuracy-psp} reports the CPU 
time  for Au ($Z=79$) to achieve $\sim 10^{-6}$ Ha accuracy across various exchange-correlation functionals, 
with all calculations again performed on an Acer Nitro AN515-46 laptop equipped with an AMD 
Ryzen 7 6800H CPU and 32~GB RAM, running Windows 11 Home (64-bit). Local and 
semilocal functionals, i.e., LDA, PBE, and rSCAN, complete in well under a second, 
while PBE0, which requires an outer-inner SCF loop structure, takes on the order of 
a few seconds due to the additional cost of exact exchange and the outer loop 
iterations. The timings for SPARC-atomSFE are comparable to those reported for the 
SPARC-atom code~\cite{bhowmik2025spectral} across the range of exchange-correlation 
functionals, despite being run on different hardware. In particular, the PBE timing of $0.21$~s for $n_{in}=8$, compared with $0.45$~s for $n_{in}=7$ in SPARC-atom  under similar hardware and with an accuracy of $\sim 10^{-6}$~Ha demonstrates the efficiency of SPARC-atomSFE.

\begin{table}[h!]
    \centering
    \begin{tabular}{@{}lccccc@{}}
        \toprule
         & $N_{fe}$ & $n_{out}$ & $n_{in}$ & Time (s) & Error (Ha) \\
        \midrule
        LDA   & $6$ & \textemdash & $7$  & $0.19$ & $5\times 10^{-6}$ \\
        PBE   & $6$ & \textemdash & $8$  & $0.21$ & $7\times 10^{-7}$ \\
        rSCAN & $3$ & \textemdash & $11$ & $0.22$ & $6\times 10^{-6}$ \\
        PBE0  & $6$ & $9$         & $53$ & $6.16$ & $7\times 10^{-7}$ \\
        \bottomrule
    \end{tabular}
    \caption{CPU time  for pseudopotential Kohn-Sham DFT calculations of 
Au ($Z=79$) for various exchange-correlation functionals, where $N_{fe}$ is chosen to 
yield the listed accuracy. $n_{out}$ and $n_{in}$ denote the number of outer and 
the total number of inner SCF iterations summed over all outer iterations, respectively.}
    \label{tab:timing-accuracy-psp}
\end{table}

\section{Concluding remarks}
\label{Sec:conclusions}

In this work, we have developed SPARC-atomSFE, a spectral finite-element package for  accurate and efficient atomic structure calculations within the framework of Kohn-Sham DFT. The package supports both all-electron and norm-conserving pseudopotential calculations and spans a comprehensive hierarchy of exchange-correlation approximations, ranging from local and semilocal to nonlocal functionals. The latter includes hybrid functionals and the many-body RPA, for which we  have implemented  both the generalized Kohn-Sham approach and the OEP method, with OEP necessary for eigenvalue-dependent functionals. Support for fractional orbital occupations and charged atoms has also been included. Spatial discretization employs an adaptive real-space grid based on the Legendre--Gauss--Lobatto scheme, high-order $C^{0}$-continuous Lagrange polynomial basis functions, and Gauss--Legendre quadrature for numerical integration. Through systematic convergence studies, we have determined the computational parameters required to achieve target accuracies. We have demonstrated the accuracy of the package through representative calculations spanning the various exchange-correlation approximations, with deviations from values in the literature generally remaining within $1~\mu\text{Ha}$.

Several natural extensions of SPARC-atomSFE suggest themselves as directions for future work. Incorporating relativistic effects through the Dirac equation would improve the fidelity of calculations for heavy elements, where such effects can be significant. Developing many-body exchange-correlation functionals beyond the RPA implementation presented here represents another promising avenue. Finally, given the growing interest in data-driven approaches to functional development, systematic integration with machine-learning workflows for training and validating exchange-correlation functionals across the full hierarchy of approximations offers another promising avenue for future research.


\section*{Acknowledgments}

The authors gratefully acknowledge the support of the U.S. Department of Energy, Office of Science, under Grant No. DE-SC0023445. The authors acknowledge useful discussions with Sayan Bhowmik (Georgia Tech.), John E Pask (LLNL) and Andrew J Medford (Georgia Tech.).


\bibliographystyle{elsarticle-num}

\end{document}